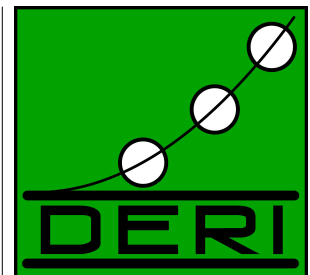

# IMPROVING THE RECALL OF DECENTRALISED LINKED DATA QUERYING THROUGH IMPLICIT KNOWLEDGE

Jürgen Umbrich      Aidan Hogan      Axel Polleres

# IMPROVING THE RECALL OF DECENTRALISED LINKED DATA QUERYING THROUGH IMPLICIT KNOWLEDGE

Jürgen Umbrich [1]        Aidan Hogan [2]        Axel Polleres [3]

**Abstract.** Aside from crawling, indexing, and querying RDF data centrally, Linked Data principles allow for processing SPARQL queries on-the-fly by dereferencing URIs. Proposed link-traversal query approaches for Linked Data have the benefits of up-to-date results and decentralised (i.e., client-side) execution, but operate on incomplete knowledge available in dereferenced documents, thus affecting recall. In this paper, we investigate how implicit knowledge – specifically that found through `owl:sameAs` and RDFS reasoning – can improve the recall in this setting. We start with an empirical analysis of a large crawl featuring 4 m Linked Data sources and 1.1 g quadruples: we (1) measure expected recall by only considering dereferenceable information, (2) measure the improvement in recall given by considering `rdfs:seeAlso` links as previous proposals did. We further propose and measure the impact of additionally considering (3) `owl:sameAs` links, and (4) applying lightweight RDFS reasoning (specifically $\rho$DF) for finding more results, relying on static schema information. We evaluate our methods for live queries over our crawl.

**Keywords:** SPARQL , Web of Data, RDFS reasoning.

---

[1]DERI, National University of Ireland, Galway, Ireland. juergen.umbrich@deri.org

[2]DERI, National University of Ireland, Galway, Ireland. aidan.hogan@deri.org

[3]DERI, National University of Ireland, Galway, Ireland / Siemens AG Österreich, Siemensstrasse 90, 1210 Vienna, Austria axel.polleres@siemens.com

**Acknowledgements**: This material is based upon works in parts jointly supported by the Science Foundation Ireland under Grant No. SFI/08/CE/I1380 (Lion-2).

# Contents

# 1  Introduction

Recently, a rich lode of RDF data has been published on the Web as *Linked Data*, by governments, academia, industry, communities and individuals alike [12]. Publishing Linked Data is governed by four principles, here summarising [1]: (**P1**) *use URIs to name things*, such that (**P2**) those *URIs can be dereferenced via HTTP*, such that (**P3**) *dereferencing yields useful RDF content* about that which is named, such that (**P4**) the returned *content includes links* (mentions external URIs) for further discovery. Given that the URIs used to name resources map (through HTTP) to the physical location of structured information about them, information published as Linked Data can be viewed as forming a scale-free, decentralised database, consisting of millions of structured Web documents [10]. Further still, thanks to the provision of typed "RDF links" between such documents [12, § 4.5], agents can traverse and navigate the resulting *Web of Data* in a manner analogous to browsing through the *Web of Documents*.

Tangentially, SPARQL [19]—the W3C standardised RDF query language—provides the declarative means to formulate structured queries against these data. Interestingly, SPARQL also encodes the notion of *Named Graphs*, which (loosely) corresponds to a means of logically partitioning some RDF corpus, such that combinations of partitions can be queried in isolation. Often, these partitions are based on the provenance of data, with the named-graph IRI corresponding to the location from which an RDF document is retrieved—*e.g.*, a Web location. Thus, given a (HTTP) correspondence between graph names and addresses, SPARQL could be supported by means of *live querying* such that the content of these graphs is retrieved from the Web at runtime. However, SPARQL semantics considers a fixed dataset from which to generate query answers, whereas live querying explores an *a priori* unbounded Web of Data for on-the-fly answers without ever considering the dataset it operates over in its entirety.

Although there is a clear symbiosis between Linked Data principles, which state that URIs should be dereferenceable, thus giving *follow-your-nose* cues as to where RDF data about a given resource might be found; and SPARQL, which gives a declarative means of stating which graphs the given query should be posed against; as of yet, SPARQL does not *formally* leverage the former set of principles. Along these lines, Hartig et al. [9] investigate using dereferenceable URIs in the query—and recursively, in the intermediate results of the query—to automatically determine a focussed set of sources which, by Linked Data principles, are likely to be *query relevant*. These query-relevant sources are then retrieved and used to generate answers to the user query, and possibly recursively, to traverse links and find further query relevant sources. When operating over sufficiently compliant Linked Data, their approach bypasses the need for source graphs to be explicitly named in the original SPARQL query and allows for new sources to be discovered in an *ad hoc* manner by traversing links at query time. Later work by Hartig [8] calls this approach *Link Traversal Based Query Execution* (LTBQE).

Note that there is an inherent trade-off in the LTBQE approach between the amount of data accessed and the recall of the response (the percentage of globally available answers returned), which varies from accessing no sources and returning no results, to (theoretically) processing the entire Web of Data. Thus, LTBQE relies on Linked Data principles as cues to identify a minimal amount of sources that maximise results.

In this paper, we first present an abstract formalism of the LTBQE approach, highlighting theoretical completeness. We then analyse the recall of the LTBQE approach in practice using an empirical analysis of a recent Linked Data crawl of ∼4 m RDF Web documents. We examine, *e.g.*, the expected percentage of dereferenceable URIs, and the ratio of triples returned in documents dereferenced to by some URI vs. all data available about that URI in the entire sample. We also look at how incorporating `rdfs:seeAlso` links into the LTBQE approach—as originally proposed by Hartig et al. [9]—affects recall. We further propose extensions of LTBQE to (i) minimise the amount of sources accessed by skipping lookups for bindings of non-distinguished variables; (ii) increase recall by considering some lightweight semantics of Linked Data which allow for (ii.a) finding additional query-relevant sources and data though consideration of `owl:sameAs` links, and (ii.b) finding additional query-relevant data through rule-based materialisation with respect to a lightweight subset of RDFS (viz. $\rho$DF [17]). With regards to (ii.b), we currently use a static set of schema data collected *a priori*. We measure the expected effect on recall for each of these extensions through analysis of our Linked Data sample. Finally, and again using our sample corpus, we randomly generate a set of benchmark queries featuring dereferenceable URIs and run them live over the remote sources, comparing all techniques discussed with respect to the number of sources accessed, the number of triples processed, and the number of answers returned.

## 2    Background and Related Work

Traditional approaches to query Linked Data locally replicate the content of the remote Linked Data sources, e.g., in a triple or quad store and execute the SPARQL queries over the local copy. Such approaches typically feature a crawler or other data acquisition component which, e.g., follows links between documents to discover new information, and/or downloads documents which have been requested for indexing by remote parties. In previous years, we supported such a service powered by YARS2 [7] allowing for querying over millions of RDF Web documents (and their entailments), but have since discontinued the endpoint due to prohibitive running costs for our research hardware. Current centralised SPARQL endpoints harvesting Linked Data include OpenLink's LOD cache[1] powered by the Virtuoso quad store [6], the FactForge [2] SPARQL endpoint[2] which includes materialised data supported by BigOWLIM [3], and more recently the Sindice [18] SPARQL endpoint[3] again powered by Virtuoso. The primary challenges for such an approach are (i) to have as much coverage of the Web of Data as possible, (ii) to keep results up to date, (iii) to be able to process potentially expressive (i.e., expensive) SPARQL queries in an efficient manner and with high concurrency. These objectives are (partially) met using distribution techniques, replication, optimised indexes, compression techniques, data synchronisation, and so on [3, 6, 7, 18]. Still, given that such services often index millions of documents, they often require large amounts of resources to run. In particular, maintaining a local, *up-to-date* index with good coverage of the Web of Data is a Sisyphean task.

Alternative approaches apply federated query processing techniques for the query execution. Recently, Ladwig and Tran [14] identify three conceptual approaches for the federated execution of SPARQL queries over Linked Data: (i) top-down query evaluation, (ii) bottom-up query evaluation, and (iii) mixed strategy query evaluation.

The top-down evaluation determines the query relevant sources before the actual query execution using knowledge about the available sources stored in a so-called "source-selection index". These source-selection indexes can vary from simple inverted-index structures [16, 18], to query-routing indexes [22], schema-level indexes [21], and lightweight hash-based structures [23].

The bottom-up query evaluation strategy involves discovering query-relevant sources on-the-fly during the evaluation of queries. The LTBQE approach [9] and the work in the present paper fall into this category. A "seed set" of remote query-relevant sources are extracted from the query; links are followed from these initial sources to find further query relevant sources and to find more answers or satisfy additional sub-goals

---

in the query; the process continues recursively until all known query-relevant sources have been exhausted. Since no local index is required, this approach can be used in decentralised scenarios, where clients can execute queries remotely over the Web without accessing a centralised service. The unique challenges for such an approach are (i) to find as many query-relevant sources as possible to improve recall of answers; (ii) to conversely minimise the amount of sources accessed to avoid traffic and slow query-response times; (iii) to optimise query execution in the absence of typical selectivity estimates, etc. [8,9]. In this paper, we focus on the first two challenges.

The third strategy involves a hybrid combination of top-down and bottom-up techniques. This strategy uses (in a top-down fashion) some knowledge about sources to map query terms or query sub-goals to sources which can contribute answers, then discovering additional query relevant sources using a bottom-up approach [14].

Such approaches translate into significant savings with respect to the resources required to locally index data; where these savings are sufficient enough, the approach can be used for decentralised querying whereby clients (those posing queries) can host the source-selection index locally on their machine. We note that there is still huge variance between the different approaches, targeting different scenarios and use-cases. For example, the inverted index traditional proposed by Sindice [18] is still very much a lightweight version of a centralised service. Conversely, our hash-based data summaries approach [23] is more geared towards lightweight, client-side processing. Depending on the particular approach taken, challenges may vary (as before) between those identified for top-down and bottom-up strategies; for example, keeping local knowledge up-to-date, or identifying a low number of query-relevant sources, etc.

A tangential approach is that of federated SPARQL querying, where remote SPARQL endpoints offer service descriptions which are indexed locally and used to route queries [20].

As such, moving towards a mature Linked Data query-answering system, one could consider a combination of approaches, where each has its complementary advantages and disadvantages. An interesting and relatively novel research area would then be investigating how to combine local and remote querying techniques both on a theoretical, engineering, and social level (cf., e.g., [10,15]). For example, using top-down approaches seems well suited for relatively static data (e.g., DBpedia, DBLP, etc.), whereas bottom-up approaches seem better suited to dynamic data (e.g., identi.ca, MusicBrainz, etc.) or potentially sensitive data, with mixed strategies falling somewhere in-between; cf [24].

In this paper, we focus on an empirical analysis of the expected recall of the LTBQE bottom-up strategy for Linked Data, as well as proposing and evaluating extensions to find additional query-relevant sources. In addition, our approach performs reasoning over the retrieved content to potentially increase recall. Our approach uses RDFS rules and a static TBox – as opposed to dynamic extraction of the TBox from fetched content as proposed by Li [16] – since we assume that terminological knowledge on the Web is relatively small (and can be kept in memory client-side [13]) and static. Moreover, we avoid "non-authoritative" redefinitions of ontology terms [13] during reasoning.

# 3   Motivating Example

Before continuing, we motivate our extensions by means of a concrete, real-world example: Figure 1 illustrates an RDF (sub)graph taken from four interlinked sources on the Web of Data.[4] The graph contains structured information about one publication (`l3sPub:HartigBF09`), "four" persons (`oh:olaf,`

---

[4]last data access on 23.06.2011

`cb:chris`, `l3sAuth:Olaf_Hartig`, `l3sAuth:Christain_Bizer`) and four dereferenceable documents.

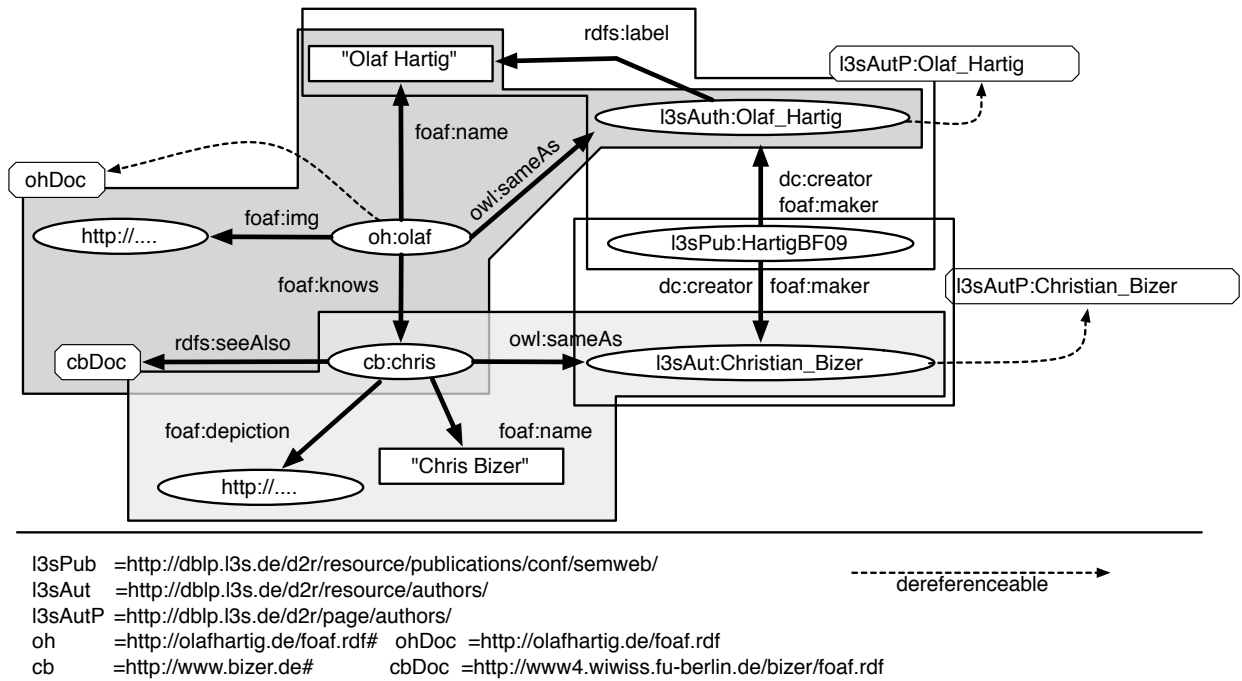

l3sPub   =http://dblp.l3s.de/d2r/resource/publications/conf/semweb/
l3sAut   =http://dblp.l3s.de/d2r/resource/authors/
l3sAutP =http://dblp.l3s.de/d2r/page/authors/
oh         =http://olafhartig.de/foaf.rdf#   ohDoc =http://olafhartig.de/foaf.rdf
cb         =http://www.bizer.de#              cbDoc =http://www4.wiwiss.fu-berlin.de/bizer/foaf.rdf

Figure 1: Snapshot of a subgraph from the Linked Open Data Web.

A typical query which can be answered by the LTQBE approach which also considers `rdfs:seeAlso` links is to ask for the pictures of friends of the person `oh:olaf` (*cf.* Query 1). The query processor evaluates this query by dereferencing the content of the query URI `oh:olaf`, following and dereferencing all URI bindings for the variable `?f` and matching the second query pattern `?f foaf:depiction ?n` over the retrieved content to find the pictures. However, the query processor needs to follow the `rdfs:seeAlso` link from `cb:chris` to `cbDoc` since the URI `cb:chris` is not dereferenceable (a dashed arrow in Fig. 1 denotes dereferenceability).

Query 2, asking for the co-authors who are also friends of `oh:olaf`, will not give results if executed with the current LTQBE approach since explicit equality information about URIs is not currently considered. The `owl:sameAs` relationship between `oh:olaf` and `l3sAuth:Olaf_Hartig` states that both URIs are equivalent and referring to the same real world entity.

Eventually, Query 3 asks for the labels of friends of `oh:olaf`. A pure link-traversal based query processor would miss the answer "Chris Bizer" because implicit knowledge encoded in ontologies used to describe real world things and their relations is not considered in the query execution. The implicit missing information to the query engine is that the `foaf:name` relation is a subProperty of the `rdfs:label` relation; particularly, note that dereferenceing the `rdfs:label` URI would not help here, since the sub-Property relation is stored in the FOAF ontology. Knowledge about the commonly used vocabularies and simple RDFS inference would address this use case.

```
SELECT ?f, ?img WHERE {
  oh:olaf foaf:knows ?f .
  ?f foaf:depiction ?img }
```
Query 1: Pictures of friends.

```
SELECT ?f WHERE {
  oh:olaf foaf:knows ?f .
  ?pub dc:creator ?f, oh:olaf}
```
Query 2: Co-authors who are friends.

```
SELECT ?o, ?l WHERE {
  oh:olaf foaf:knows ?f .
  ?f rdfs:label ?l }
```
Query 3: Labels of friends

# 4 Preliminaries

In this section, we cover some necessary preliminaries and notation relating to RDF (§ 4.1), Linked Data (§ 4.2), SPARQL (§ 4.3) and RDFS & OWL (§ 4.4).

## 4.1 RDF

We first provide some standard notation for dealing with RDF [11].

**Definition 4.1 (RDF Term, Triple and Graph)**
*The set of* RDF terms *consists of the set of URIs* $\mathbf{U}$, *the set of blank-nodes* $\mathbf{B}$ *and the set of literals* $\mathbf{L}$ *(which includes plain and datatype literals). An* RDF triple $t := (s, p, o)$ *is an element of the set* $\mathbf{G} := \mathbf{UB} \times \mathbf{U} \times \mathbf{UBL}$ *(where, e.g.,* $\mathbf{UB}$ *is a shortcut for set-union). Here $s$ is called subject, $p$ predicate, and $o$ object. A finite set of RDF triples $G \subset \mathbf{G}$ is called an* RDF graph. *We use the functions* $\mathsf{subj}(G)$, $\mathsf{pred}(G)$, $\mathsf{obj}(G)$, $\mathsf{terms}(G)$, *to denote the set of all terms projected from the subject, predicate, object and any position of a triple $t \in G$ respectively.*

## 4.2 Linked Data

The Linked Data principles [1] have already been introduced in § 1. We now provide some notation which helps to formalise these principles and relate them to RDF. As per [9], we currently do not consider temporal issues with, e.g, HTTP-level functions.

**Definition 4.2 (Data Source and Linked Dataset)**
*We define the* http-download *function* $\mathsf{get} : \mathbf{U} \to 2^{\mathbf{G}}$ *as the mapping from URIs to RDF graphs provided by means of HTTP lookups which directly return status code* 200 OK *and data in a suitable RDF format. We define the set of (RDF)* data sources $\mathbf{S} \subset \mathbf{U}$ *as the set of URIs* $\mathbf{S} := \{s \in \mathbf{U} : \mathsf{get}(s) \neq \emptyset\}$. *We define a Linked Dataset as* $\Gamma \subset \mathsf{get}$; *i.e., a finite set of pairs* $(u, \mathsf{get}(u))$, *and* $\mathsf{merge}(\Gamma) := \biguplus_{(u,G) \in \Gamma} G$ *as the RDF merge of graphs in $\Gamma$ which preserves the uniqueness of blank-node labels across graphs [11].*

**Definition 4.3 (Dereferencing RDF)**   *A URI may issue a HTTP redirect to another URI with a* 30x *response code; we denote this function as* $\mathsf{redir} : \mathbf{U} \to \mathbf{U}$ *which maps a URI to itself in the case of failure (e.g., where no redirect exists); note that* $\mathsf{redir}$ *would* also *strip the fragment identifier of a URI if present. We denote the fixpoint of* $\mathsf{redir}$ *as* $\mathsf{redirs}$, *denoting traversal of a number of redirects (a limit may be imposed to avoid cycles). We denote* dereferencing *by the composition* $\mathsf{deref} := \mathsf{get} \circ \mathsf{redirs}$ *which maps a URI to an RDF graph retrieved with status code* 200 OK *after following redirects, or which maps a URI to the empty set in the case of failure. We denote the set of* dereferenceable URIs *as* $\mathbf{D} := \{d \in \mathbf{U} : \mathsf{deref}(d) \neq \emptyset\}$; *note that* $\mathbf{D} \subset \mathbf{S}$ *and we place no expectations on what* $\mathsf{deref}(d)$ *returns (as long as it returns* some *valid RDF).*

### 4.3 SPARQL

We now introduce some concepts relating to SPARQL [19]. Note that herein, we focus on evaluating simple, conjunctive, *basic graph patterns* (BGPs), where we do not consider do not consider more expressive parts of the SPARQL language (although they can be layered on top of our methods).

**Definition 4.4 (Variables, Triple Patterns and Queries (BGPs))**
*Let* $\mathbf{V}$ *be the set of variables ranging over* $\mathbf{UBL}$. *A triple pattern* $tp := (s, p, o)$ *is an element of the set* $\mathbf{Q} := \mathbf{VUL} \times \mathbf{VU} \times \mathbf{VUL}$. *For simplicity, we do not consider blank-nodes in triple patterns (they could be roughly emulated by an injective mapping from* $\mathbf{B}$ *to* $\mathbf{V}$*). A finite (herein, non-empty) set of triple patterns* $Q \subset \mathbf{Q}$ *is called a* Basic Graph Pattern*, or herein, simply a* query*. We use* $\mathsf{vars}(Q) \subset \mathbf{V}$ *to denote the set of variables in Q. Finally, we may overload graph notation for queries, where, e.g.,* $\mathsf{terms}(Q)$ *returns all elements of* $\mathbf{VUL}$ *in Q.*

**Definition 4.5 (SPARQL solutions)**
*Call the partial function* $\mu : \mathrm{dom}(\mu) \cup \mathbf{UL} \to \mathbf{UBL}$ *a* solution mapping *which grounds variables in* $\mathrm{dom}(\mu) \subset \mathbf{V}$ *to* $\mathbf{UBL}$ *and which is the identify function for* $\mathbf{UL}$. *Overloading notation, let* $\mu : \mathbf{Q} \to \mathbf{G}$ *and* $\mu : 2^{\mathbf{Q}} \to 2^{\mathbf{G}}$ *also resp. denote a solution mapping from triple patterns to RDF triples, and basic graph patterns to RDF graphs such that* $\mu(tp) := (\mu(s), \mu(p), \mu(o))$ *and* $\mu(Q) := \{\mu(tp) \mid tp \in Q\}$. *Now, we define the set of solutions for a query Q over a Linked Dataset* $\Gamma$ *as* $\Omega(\Gamma, Q) := \{\mu \mid \mu(Q) \subseteq \mathsf{merge}(\Gamma) \wedge \mathrm{dom}(\mu) = \mathsf{vars}(Q)\}$. *Note that herein, and unlike SPARQL, solutions are given as sets (not multi-sets), implying a default* `DISTINCT` *semantics for queries.*

### 4.4 RDFS and OWL

We define some preliminaries relating to RDFS and OWL. In particular, we support a subset of OWL 2 RL/RDF rules, given in Table 1, which constitute a partial axiomatisation of the OWL RDF-Based Semantics. Our RDFS rules are a subset of the $\rho$DF rules proposed by Muñoz et al. [17] which deal with instance data entailments (as opposed to schema-level entailments).[5] Our subset of OWL rules are specifically chosen to support the semantics of equality (particularly replacement) for `owl:sameAs`. Note that these rules support the RDFS/OWL features originally recommended for use by Bizer et al. when publishing Linked Data [4, §4.2, §6]. The rules we consider are given in Table 1. More recent guidelines [12, §4.4.3] recommend use of additional OWL features; we leave support for more expressive OWL reasoning to future work. For convenience, we re-use previous notation in the following formalisms.

**Definition 4.6 (Entailment Rules and Least Model)**
*An entailment rule is a pair* $r = (Body, Head)$ *such that* $Body, Head \subset \mathbf{Q}$*; and* $\mathsf{vars}(Head) \subseteq \mathsf{vars}(Body)$. *The immediate consequences of r for a Linked Dataset* $\Gamma$ *are denoted and given as:*

$$\mathfrak{T}(\Gamma, r) := \{\mu(Head) \mid \mu \in \Omega(\Gamma, Body)\}$$

*Let R denote a set of entailment rules. The immediate consequences of R over* $\Gamma$ *are given as* $\mathfrak{T}(\Gamma, R) := \bigcup_{r \in R} \mathfrak{T}(\Gamma, r)$. *Further, let* $v$ *denote a fresh URI which names the graph G of inferred data, let* $G_0 = \emptyset$, $^R\mathfrak{T}_0 := \Gamma \cup \{(v, G_0)\}$ *and let* $^R\mathfrak{T}_i := \Gamma \cup (\{v, \mathfrak{T}(^R\mathfrak{T}_{i-1}, R) \cup G_{i-1})\}$ *for* $i \in \mathbb{N}$. *The* least model *of R for* $\Gamma$ *is given as the least n such that* $^R\mathfrak{T}_n = {}^R\mathfrak{T}_{n+1}$ *(i.e., the closure is reached); we denote the least model as* $^R\mathfrak{T}$. *Query answers incl. entailments are given by* $\Omega(^R\mathfrak{T}, Q)$.

---

[5]We drop *implicit typing* [17] rules as we allow generalised RDF in intermediate inferences.

| ID | Body | | Head |
|---|---|---|---|
| PRP-SPO1 | ?p$_1$ rdfs:subPropertyOf ?p$_2$ . | ?s ?p$_1$ ?o . | ?s ?p$_2$ ?o . |
| PRP-DOM | ?p rdfs:domain ?c . | ?s ?p ?o . | ?p a ?c . |
| PRP-RNG | ?p rdfs:range ?c . | ?s ?p ?o . | ?o a ?c . |
| CAX-SCO | ?c$_1$ rdfs:subClassOf ?c$_2$ . | ?s a ?c$_1$ . | ?s a ?c$_2$ . |
| EQ-SYM | ?x owl:sameAs ?y . | | ?y owl:sameAs ?x . |
| EQ-TRANS | ?x owl:sameAs ?y . | ?y owl:sameAs ?z . | ?x owl:sameAs ?z . |
| EQ-REP-S | ?s owl:sameAs ?s$'$ . | ?s ?p ?o . | ?s$'$ ?p ?o . |
| EQ-REP-P | ?p owl:sameAs ?p$'$ . | ?s ?p ?o . | ?s ?p$'$ ?o . |
| EQ-REP-O | ?o owl:sameAs ?o$'$ . | ?s ?p ?o . | ?s ?p ?o$'$ . |

Table 1: $\rho$DF and `owl:sameAs` rules with OWL 2 RL/RDF naming

## 5 Link Traversal Based Query Execution

We now discuss the Link Traversal Based Query Execution (LTBQE) approach introduced by Hartig et al. [9], where we present a novel, abstract formalisation and characterise the completeness of the approach with respect to the Web of Data (§ 5.1). We then look at extensions of the approach which improve recall (§ 5.2).

### 5.1 Baseline LTBQE

**Definition 5.1 (LTBQE Query Relevant Sources and Answers)**   *Define* derefs $: 2^{\mathbf{U}} \rightarrow \mathbf{U} \times 2^{\mathbf{G}}$; $U \mapsto \{(\text{redirs}(u), \text{deref}(u)) \mid u \in U)\}$ *as the mapping from a set of URIs to the Linked Dataset it represents by dereferencing all URIs. Given a BGP query $Q$ as before, let $U_Q :=$ terms$(Q) \cap \mathbf{U}$ denote the set of URIs appearing in $Q$. Let $\Gamma_0^Q :=$ derefs$(U_Q)$ represent the dataset retrieved by dereferencing all query URIs.[6] Next let* uris$(\mu) := \{u \in \mathbf{U} \mid \exists v \text{ s.t. } (v, u) \in \mu\}$ *denote the set of URIs in a solution mapping $\mu$, and let* $U_i := \{u \in \text{uris}(\mu) \mid \exists \mu, \exists tp \in Q \text{ s.t. } \mu(\{tp\}) \subseteq \text{merge}(\Gamma_{i-1}^Q)\}$ *for $i \in \mathbb{N}$ be the set of URIs which appear as a solution mapping for a triple pattern in $Q$ for the dataset $\Gamma_{i-1}^Q$, and let $\Gamma_i^Q :=$ derefs$(U_i) \cup \Gamma_0^Q$.[7] The set of* LTBQE query relevant sources *for $Q$ is given as the least $n$ such that $\Gamma_n^Q = \Gamma_{n+1}^Q$, denoted simply $\Gamma^Q$. The set of* LTBQE query answers *for $Q$ is given as $\Omega(\Gamma^Q, Q)$, or simply $\Omega^Q$.*

With regards to completeness, let get denote the dataset (theoretically) represented by the entire Web of Data (note: get $\subset \mathbf{U} \times 2^G$). One may then ask when $\Omega^Q$ is complete with respect to get. A trivial sufficient condition for completeness is given by $\Gamma^Q =$ get. A "top down" condition is given by looking at all of the answers available for get:

**Proposition 5.1**  *Let $Q$ be a query s.t.* terms$(Q) \cap \mathbf{U} \neq \emptyset$. *Let $\Omega(\text{get}, Q)$ denote the set of answers given by the Web of Data. For each $\mu \in \Omega(\text{get}, Q)$, let $D := (V, A)$ denote a directed graph with vertices $V = \mathbf{U} \cap$ terms$(\mu(Q))$ and arcs:*

$$E := \{(x, y) \in V \times V \mid \exists t \in \mu(Q) \text{ s.t. } \{x, y\} \subseteq \text{terms}(\{t\}) \text{ and } t \in \text{deref}(s)\}$$

*Again denote the query URIs by $U_Q :=$ terms$(Q) \cap \mathbf{U}$, let $A^+$ denote the transitive closure of the arcs, and let*

$$U_Q^+ := U_Q \cup \{u \in \mathbf{U} \mid \exists u_Q \in U_Q \text{ s.t. } (u_Q, u) \in A^+\}$$

---

[6]One could consider $\Gamma_0^Q$ as also containing "seed" data [9].

[7]Or, equivalently (for static data) $\Gamma_i^Q := \Gamma_{i-1}^Q \cup \text{derefs}(U_i \setminus U_{i-1})$.

*denote the set of URIs reachable in the graph from a query URI. Now, $\Omega^Q$ contains a solution isomorphic to $\mu$ (up to blank-node relabelling) if and only if $\mathsf{merge}(\mathsf{derefs}(U_Q^+)) \models \mu(Q)$ (i.e., the former simple-entails the latter [11]).*

**Proof 5.1** *(Sketch) The above proposition essentially rephrases the LTBQE algorithm to highlight the reachability condition from a query URI, where $\Gamma^Q$ is equivalent to $\mathsf{derefs}(U_Q^+)$.*

The proposition has some interesting consequences. First, given a query with no dereferenceable URIs, LTBQE cannot return results. Second, given a query with multiple URIs, different reachability conditions can occur from different starting points; thus, all query URIs must be initially retrieved [9]. Third, answers "connected" by literals or involving blank-nodes in unreachable documents will often affect completeness. Fourth, in the general case, reachability is heavily dependent on the amount of data returned by the $\mathsf{deref}(u)$ function, which would ideally return all triples mentioning $u$ on the Web of Data. The fourth assumption is clearly idealised; hence, in Section 6 we will empirically analyse how much the assumption holds in practice, giving insights into the recall of LTBQE. Beforehand, we propose some extensions to extend LTBQE recall by considering implicit knowledge.

### 5.2 Extending LTBQE

We now look at three extensions to the baseline LTBQE approach.

**1. Following `rdfs:seeAlso`:** The first extension is that proposed by Hartig et al., which considers `rdfs:seeAlso` links to extend the set of query-relevant sources. Here, we briefly summarise. Adapting Definition 5.1, let $\bar{\Gamma}_0^Q := \Gamma_0^Q$ and let:

$$\bar{U}_i := U_i \cup \{u \in \mathbf{U} \mid \exists u' \in U_i \text{ s.t. } (u', \texttt{rdfs:seeAlso}, u) \in \mathsf{merge}(\bar{\Gamma}_{i-1}^Q)\},$$

let $\bar{\Gamma}_i^Q := \mathsf{derefs}(\bar{U}_i) \cup \Gamma_0^Q$, and finally let $\bar{\Gamma}^Q$ be the fixpoint as before and let $\bar{\Omega}^Q$ be the respective solutions. Here, `rdfs:seeAlso` links are used to expand the set of query relevant sources.

**2. Following and reasoning over `owl:sameAs`:** We propose an extension of LTBQE to consider `owl:sameAs` inferences. Let $R$ denote the set of rules of the form EQ-* in Table 1. Let now $\mathbb{T}_0^Q := {}^R\mathbb{T}_0^Q$ (recalling ${}^R\mathbb{T}$ from Def. 4.6 and $\Gamma_0^Q$ from Def. 5.1), and let:

$$U_i' := \{u \in \mathsf{uris}(\mu) \mid \exists \mu, \exists tp \in Q \text{ s.t. } \mu(\{tp\}) \subseteq \mathsf{merge}(\mathbb{T}_{i-1}^Q)\},$$

$${}^{\mathcal{C}}U_i := \{u \in \mathbf{U} \mid \exists u' \in U_i' \text{ s.t. } (u', \texttt{owl:sameAs}, u) \in \mathsf{merge}(\mathbb{T}_{i-1}^Q)\},$$

where $\mathbb{T}_i^Q := {}^R\mathsf{derefs}({}^{\mathcal{C}}U_i) \cup \mathbb{T}_0^Q$, and finally let $\mathbb{T}^Q$ be the fixpoint as before and let ${}^{\mathcal{C}}\Omega^Q$ be the respective solutions. Here, `owl:sameAs` links are used to expand the set of query relevant sources, and `owl:sameAs` rules are used to materialise implicit knowledge given by the OWL semantics, potentially generating additional answers.

**3. Reasoning for $\rho$DF:** We propose a final novel extension of LTBQE to consider a subset of RDFS reasoning as per the `prp-*` and `cax-sco` rules in Table 1, which we again denote here by $R$. We currently consider a static set of schema data representing vocabularies on the Web, which we denote by $\Gamma^{voc}$. This serves as input into the LTBQE algorithm. In future work, we plan to investigate dereferencing schema knowledge live from the Web of Data.

Now, adapting Definition 5.1, let $_\rho\Gamma_0^Q := \Gamma^{voc} \cup {}^R\Gamma_0^Q$ and let:

$$_\rho U_i := \{u \in \mathsf{uris}(\mu) \mid \exists\mu, \exists tp \in Q \text{ s.t. } \mu(\{tp\}) \subseteq \mathsf{merge}(_\rho\Gamma_{i-1}^Q)\}\,,$$

where $_\rho\Gamma_i^Q := {}^R\mathsf{derefs}(_\rho U_i) \cup {}_\rho\Gamma_0^Q$, and finally let $_\rho\Gamma^Q$ be the fixpoint as before and let $_\rho\Omega^Q$ be the respective solutions. Here, RDFS rules and background schema knowledge are used to materialise implicit knowledge, potentially generating additional answers (and thus possibly finding new query relevant sources).

**Combined** Of course, the above methods can be combined in a natural fashion, where, e.g., for combining all extensions, the query relevant sources are denoted $_\rho^e\bar{\Gamma}^Q$ and the answers by $_\rho^e\bar{\Omega}^Q$.

# 6 Empirical Study

In Section 5.1, we mentioned that the recall of the LTBQE approach is—in the general case—dependent on the dereferenceability of data. Along those lines, we now present the results of our empirical study of a Linked Data corpus. We survey the ratio of all triples mentioning a URI in our corpus against those returned in the dereferenceable document of that URI; we do so for different triple positions. We also look at the comparative recall of data considering (1) explicit, dereferenceable information; (2) including `rdfs:seeAlso` links [9]; (3) including `owl:sameAs` links and implicit knowledge; (4) including RDFS reasoning.

**Empirical corpus** Our corpus was crawled in mid-May 2010 from 3.985 million RDF/XML documents spanning 783 pay-level domains (data providers). URIs in all triples positions were considered for crawling. We conducted the crawl in a breadth-first manner to guarantee a broad sample of data-providers (and to ensure polite crawling), we assign each pay-level domain (PLD) an individual priority queue. The PLD queues are sampled in a round-robin fashion during the crawl, with the highest linked URIs *for each domain* being returned first.

The resulting corpus contains 1.118 g quadruples (965 m unique triples) mentioning 286 m RDF terms, of which 29 m (10%) are Literals, 165 m (57.8%) are blank nodes, and 92 m (32.2%) are URIs. We denote the corpus as $\Gamma_\sim$. It's important to note that this corpus is only a *sample* of the Web of Data; in particular, we have not performed a HTTP lookup on all URIs in the corpus, where we looked up—and thus only have knowledge of redir and deref functions for—8.63 m URIs; all of these URIs are HTTP and do not have non-RDF file-extensions. We denote these URIs by $U_\sim$. Of the 8.63 m, 5.65 m (65.5%) dereferenced to RDF; we denote these by $D_\sim$. Further note that, wrt. the Web of Data, our sample recall measures specify an upper bound.

*More details about the parameters of the crawl and various statistics about the corpus are presented in [13].*

**RDFS Schema**    From our corpus, we extract a static set of schema data for the RDFS reasoning. As argued in [5], schema data on the Web is often noisy, where third-party publishers "redefine" popular terms outside of their namespace; for example, one document defines nine *properties* as the domain of `rdf:type`, which would have a drastic effect on our reasoning.[8] Thus, we perform authoritative reasoning, which conservatively discards certain third-party schema axioms (cf. [5]). Thus, our schema data only considers triples of the following form:

$$(s, \mathtt{rdfs:subPropertyOf}, o) \in \mathsf{deref}(s), (s, \mathtt{rdfs:subClassOf}, o) \in \mathsf{deref}(s)$$
$$(s, \mathtt{rdfs:domain}, o) \in \mathsf{deref}(s), (s, \mathtt{rdfs:range}, o) \in \mathsf{deref}(s)$$

We extracted 397,749 such authoritative RDFS triples from 70 PLDs as follows: 229,575 `rdfs:subClassOf` (55 PLDs); 12,266 `rdfs:subPropertyOf` (43 PLDs); 85,071 `rdfs:domain` (57 PLDs); 70,837 `rdfs:range` (58 PLDs).

## 6.1   Recall for Baseline

We first measure the average dereferenceability of information in our sample. For a dereferenceable uri $d$, we compute the *sample dereferencing recall* $\mathsf{sdr}(d)$ as the ratio of the number of unique triples mentioning $d$ in $\mathsf{deref}(d)$ vs. unique triples mentioning $d$ across the entire sample. We denote by $\mathsf{sdr}_\sim$ the average $\mathsf{sdr}(d)$ for all $d \in D_\sim$. We also analyse the $\mathsf{sdr}(d)$ restricting the specific triple positions where $d$ appears. We ignore $d$ in the average if it does not appear in the relevant triple position in the sample.

Table 2 presents the results, where for different triples positions we present (i) number of URIs appearing in that position, (ii) number of which were dereferenceable, (iii) ratio of dereferenceable URIs, (iv) $\mathsf{sdr}_\sim$ for that position, (v) standard deviation of $\mathsf{sdr}_\sim$. Column *type*-**object** considers $d$ appearing as object in an `rdf:type` triple separately.

The analysis provides some interesting insights into the LTBQE approach. Given a HTTP URI without a common non-RDF extension, we have a 65% success ratio to receive RDF/XML content regardless of the triple position; for subjects, the percentage increases to 84%, etc. If such a URI dereferences to RDF, we receive on average 65% of all triples in which it appears. Given a triple pattern with a URI in the subject position, the dereferenceable ratio increases to 95%; for objects, the ratio drops to 44%; LTBQE would perform poorly for triple patterns with (only) a URI in the predicate position; etc. In summary, LTBQE performs well when URIs appear in the subject position of triple patterns, moderately when URIs appear in the object, but poorly when URIs appear in the predicate or object of an `rdf:type` triple.

| | **any** | **subject** | **predicate** | **object** | *type*-**object** |
|---|---|---|---|---|---|
| $|U_\sim|$ | 8.63 m | 6.51 m | 15.15 k | 6.60 m | 74.72 k |
| $|D_\sim|$ | 5.65 m | 5.46 m | 635 | 3.02 m | 21.2 k |
| $|U_\sim|/|D_\sim|$ | 0.65 | 0.84 | 0.04 | 0.46 | 0.28 |
| `AVG` $\mathsf{sdr}_\sim$ | **0.65** | **0.95** | **0.00009** | **0.44** | **0.0037** |
| `SD` $\mathsf{sdr}_\sim$ | 0.47 | ±0.19 | ±0.009 | ±0.47 | ±0.06 |

Table 2: Dereferenceability results for different triple positions

---
[8]viz. `http://www.eiao.net/rdf/1.0`

## 6.2  Recall for Extensions

We now measure the `sdr` increase given by extending LTBQE to also consider `rdfs:seeAlso` and `owl:sameAs` links, as well as implicit knowledge given by `owl:sameAs` and RDFS reasoning.

**Benefit of following `rdfs:seeAlso` links**   We measured the percentage of dereferenceable URIs in $D_\sim$ which have at least one `rdfs:seeAlso` link in their dereferenced document to be 7% for our sample. Where such links exist, following them increases the amount of unique triples by a factor of $1.016\times$ vs. the unique triples in the dereferenced document alone. We conclude that, in the general case, considering `rdfs:seeAlso` information for the query processing will only marginally effect the recall increase of LTBQE.

**Benefit of following `owl:sameAs` links & implicit knowledge**   We measured the percentage of dereferenceable URIs in $D_\sim$ which have at least one `owl:sameAs` links in their dereferenced document to be 18% for our sample. Where such links exist, following them and applying the `EQ-*` entailment rules over the resulting information increases the amount of unique triples by a factor of $3.85\times$ vs. the unique (explicit) triples in the dereferenced document alone. We conclude that, in the general case, `owl:sameAs` links are not so commonly found for dereferenceable URIs, but where available, following them and applying the entailment rules generates significantly more data for generating answers.

**Benefit of including $\rho$DF implicit knowledge**   We measured the percentage of dereferenceable URIs in $D_\sim$ whose dereferenced documents given non-empty unique entailments through authoritative $\rho$DF reasoning with respect to $\Gamma^{voc}$ as 98%. Where such entailments are non-empty, they increase the amount of unique triples by a factor of $2.44\times$ vs. the unique (explicit) triples in the dereferenced document. We conclude that such reasoning often increases the amount of data available for LTBQE query answering, and by a significant amount.

# 7  Evaluation

We evaluate our proposed LTBQE extensions for a set of pseudo-randomly generated queries over our Linked Data corpus. Our main goal is to compare and contrast different setups and to assess our proposed extensions in a realistic scenario.

**Implementation**   We have implemented Hartig et al.'s iterator-based algorithm for LTBQE (which was shown to be complete) [9]. We use ARQ to parse and process input SPARQL queries.[9]   We further use the LDSpider crawling framework for performing live Linked Data lookups; LDSpider respects the `robots.txt` policy, blacklists typical non-RDF URI patterns (*e.g.,* `.jpeg`) and enforces a half-second delay between two consequential lookups for URIs hosted at the same domain.[10] We use the SAOR engine to support the aforementioned rule-based reasoning extensions [5]. Note that we use the same input RDFS data as used in the empirical study of the previous section.

---

[9]`http://jena.sourceforge.net/ARQ/`
[10]`http://code.google.com/p/ldspider/`

**Optimised LTBQE**  Inspired by our empirical analysis, we also implement and evaluate a variation of the LTBQE approach which does not dereference URIs appearing only in the predicate position of a (possibly partially bound) triple-pattern. Further, we add another optimisation which avoids dereferencing URIs which are only bound by non-distinguished variables not appearing elsewhere in the query (i.e., variables whose value is not used elsewhere). Since these optimisation reduce the number of query-relevant sources, they may in theory lead to less results, though in practice (and as per our empirical survey), we would expect a minimal change in recall over the baseline.

## 7.1   Query Generation

**Query shapes and generation**  We benchmark queries of elemental graph shapes, *viz.,* entity, star and path queries.

**Entity queries** (*entity*-[*s*|*o*|*so*]) ask for all available triples for an entity. We generate three types of entity queries, asking for triples where a URI appears (*entity-s*) as the subject; (*entity-o*) as the object; (*entity-so*) as the subject *or* object. An example for *entity-so* would be {`<d> ?p1 ?o .  ?s ?p2 <d> .`}

**Star queries** (*star*-[*s3*|*o3*|*s1-o1*|*s2-o1*|*s1-o2*]) contain three acyclic triple patterns which share exactly one URI (called the center node) and where predicate terms are constant. We generate four different variations of such queries, differing in the number of triple patterns in which the center node appears at the subject (s) or object (o). Thus, each query has 4 constants and 3 variables. An example for *star-s2-o1* would be {`<d> foaf:knows ?o ;foaf:name ?o1 .  ?o3 dc:creator <d> .`}

**Path queries** ([*s*|*o*]-*path*-[*2*|*3*]) consist of 2 or 3 triple patterns which form a path—precisely two triple pattern share the same variable. Exactly one triple pattern has a URI at either the subject or object position and all predicate terms are constant. We generate four different sub-types: path shaped queries of length 2 and 3 in which either the subject or object term of one of the triple pattern is a constant. An example for *s-path-2* is Query 1.

In total, we generate 200 `SELECT DISTINCT` queries for *each* of the above 12 query shapes using random walks in our corpus. To help ensure that queries return non-empty results (in case there are no HTTP connection errors or time outs) we consider dereferenceable information for the query generation which (1) picks randomly a pay-level-domain available in the dereferenceable URIs $D_\sim$, (2) selects randomly a URI from $D_\sim$ for that PLD and (3) generates appropriate triple patterns from the dereferenceable document of the selected URI. For path shape queries, when performing steps (2) and (3), the URI for the next triple pattern is selected out of the URIs contained for the previous triple pattern, as per a random walk of dereferenceable URIs.

Distinguished variables are picked by randomly choosing a single variable as distinguished and make further variables distinguished with a probability of 0.5.

## 7.2   Results

We measure for each query the following six metrics: (1) the number of distinct results (**result**), (2) the total time to execute the query (**time**), (3) time elapsed until the first result was returned (**first**), (4) number of total lookups (**http**), (5) total number of triples processed (**retrieved**) and (6) total number of inferred triples (**inferred**).

We execute each query with six different setups: `base` gives the LTBQE baseline; `select` denotes optimised LTBQE; `seeAlso`, `sameAs`, and $\rho$DF extend optimised LTBQE; `combined` denotes all extensions over optimised LTBQE. Note that we present wall-clock times "as-is": running queries live over the Web of Data introduces many external variables which we cannot account for.

We ran our 200 benchmark queries directly over the Web of Data. We encountered various HTTP-level issues which affected comparability of results for the six setups. Thus, we select and present results for "stable" queries which returned a non-empty results and where none of the required HTTP lookups resulted in a response code of *5xx* or other connection errors (e.g., timeouts). The detailed benchmark results are listed in Table 3. Overall, the expected recall improvements from our empirical analyses for the different optimisations are verified by our experiments. Our source selection optimisation reduces the number of total lookups by two to three times without significantly affecting the result recall. We observed that considering `rdfs:seeAlso` information for the query processing marginally affects the content recall but has no measurable influence on the results. Further, we observe that following `owl:sameAs` links, if available, and applying the entailment rules generates significantly more data and increase the number of returned answers. A similar increase in data and results can be measured if we consider $\rho$DF inferences. The combination of all optimisation returned for all query the most answers and effectively increased the amount of query relevant data.

## 8   Conclusion

Proposed link-traversal query approaches for Linked Data have the benefit of up-to-date results and decentralised execution, but operate over incomplete knowledge available in dereferenced documents, thus affecting recall for results. We empirically study this issue for a large sampling of the Web of Data, consisting of 4 m Linked Data sources and 1.1 g quadruples. We further propose to improve recall by considering implicit knowledge, specifically that found through `owl:sameAs` and RDFS reasoning. We again validate our extensions by analysis of our corpus, where we show increases in data available to the LTBQE approach (1) of $1.02\times$ considering `rdfs:seeAlso` information as proposed in [9], (2) of $3.8\times$ considering `owl:sameAs` and (3) of $2.4\times$ if we apply $\rho$DF reasoning using static schema information. We further generate and run queries (of twelve different shapes) live over the Web of Data, comparing six different setups, and demonstrating improvements for our extensions in the average number of answers generated, albeit at the cost of accessing more data.

Our source code and stable experimental queries are available at `http://code.google.com/p/lidaq/wiki/Lidaq`.

*Future Work* We plan to extend our entailment rules to cover more of OWL (2) and investigating changes in recall when considering dynamically dereferenced schema data vs. static schema data. We also plan to investigate `owl:sameAs` optimisations for canonicalising equivalent URIs as opposed to materialising all equivalent data.

## References

[1] T. Berners-Lee. Linked Data. Design issues, W3C, 2006.

[2] B. Bishop, A. Kiryakov, D. Ognyanoff, I. Peikov, Z. Tashev, R. Velkov. FactForge: A fast track to the web of data. *Sem. Web J.*, (to appear), 2011.

[3] B. Bishop, A. Kiryakov, D. Ognyanoff, I. Peikov, Z. Tashev, R. Velkov. OWLIM: A family of scalable semantic repositories. *SWJ*, 2(1):33–42, 2011.

[4] C. Bizer, R. Cyganiak, T. Heath. How to Publish Linked Data on the Web. `linkeddata.org` Tutorial, July 2008.

[5] P. A. Bonatti, A. Hogan, A. Polleres, L. Sauro. Robust and Scalable Linked Data Reasoning Incorporating Provenance and Trust Annotations. *JWS*, 2011.

[6] O. Erling, I. Mikhailov. RDF Support in the Virtuoso DBMS. *Networked Knowledge – Networked Media*, volume 221, p. 7–24. Springer, 2009.

[7] A. Harth, J. Umbrich, A. Hogan, S. Decker. YARS2: A Federated Repository for Querying Graph Structured Data from the Web. *ISWC*, 2007.

[8] O. Hartig. Zero-knowledge query planning for an iterator implementation of link traversal based query execution. *ESWC (1)*, p.154–169, 2011.

[9] O. Hartig, C. Bizer, J.-C. Freytag. Executing sparql queries over the web of linked data. *ISWC'09*, 2009.

[10] O. Hartig, A. Langegger. A database perspective on consuming linked data on the web. *Datenbank-Spektrum*, 10(2):57–66, 2010.

[11] P. Hayes. RDF Semantics. W3C Recommendation, Feb. 2004.

[12] T. Heath, C. Bizer. *Linked Data: Evolving the Web into a Global Data Space*, volume 1. Morgan & Claypool, 2011.

[13] A. Hogan. *Exploiting RDFS and OWL for Integrating Heterogeneous, Large-Scale, Linked Data Corpora*. PhD thesis, DERI, NUIG, 2011.

[14] G. Ladwig, T. Tran. Linked data query processing strategies. *ISWC (1)*, p. 453–469, 2010.

[15] G. Ladwig, T. Tran. Sihjoin: Querying remote and local linked data. *ESWC (1)*, p. 139–153, 2011.

[16] Y. Li, J. Heflin. Using reformulation trees to optimize queries over distributed heterogeneous sources. *ISWC*, p. 502–517, 2010.

[17] S. Muñoz, J. Pérez, C. Gutierrez. Simple and Efficient Minimal RDFS. *J. Web Sem.*, 7(3):220–234, 2009.

[18] E. Oren, R. Delbru, M. Catasta, R. Cyganiak, H. Stenzhorn, G. Tummarello. Sindice.com: a document-oriented lookup index for open linked data. *IJMSO*, 3(1):37–52, 2008.

[19] E. Prud'hommeaux, A. Seaborne. SPARQL Query Language for RDF. W3C Recommendation, Jan. 2008. `http://www.w3.org/TR/rdf-sparql-query/`.

[20] B. Quilitz, U. Leser. Querying distributed rdf data sources with sparql. *ESWC*, p. 524–538, 2008.

[21] H. Stuckenschmidt, R. Vdovjak, G.-J. Houben, J. Broekstra. Index structures and algorithms for querying distributed RDF repositories. *WWW*, 2004.

[22] T. Tran, L. Z. 0007, R. Studer. Summary models for routing keywords to linked data sources. *ISWC (1)*, p. 781–797, 2010.

[23]  J. Umbrich, K. Hose, M. Karnstedt, A. Harth, A. Polleres. Comparing data summaries for processing live queries over linked data. *WWWJ*, p. 1–50, 2011.

[24]  J. Umbrich, B. Villazón-Terrazas, M. Hausenblas. Dataset Dynamics Compendium: A Comparative Study. *COLD Workshop*, 2010.

| Setup | Results | | Time (sec) | | First (sec) | | HTTP | | Retrieved (k) | | Inferred (k) | |
|---|---|---|---|---|---|---|---|---|---|---|---|---|
| **Query class entity-s with 79 queries** | | | | | | | | | | | | |
| base | 10.68 | (±10.2) | 7.97 | (±11.52) | 1.02 | (±1.98) | 17.72 | (±16.95) | 3.41 | (±10.59) | 0 | (±0) |
| select | 10.67 | (±10.2) | 4.08 | (±11.81) | 1.49 | (±6.55) | 5.33 | (±11.93) | 2.26 | (±10.30) | 0 | (±0) |
| seeAlso | 10.67 | (±10.2) | 3.36 | (±7.66) | 1.03 | (±2.53) | 5.52 | (±11.97) | 2.26 | (±10.30) | 0 | (±0) |
| sameAs | 14.72 | (±24.98) | 7.27 | (±25.92) | 0.88 | (±1.5) | 13.96 | (±97.99) | 10.81 | (±40.36) | 8.16 | (±45.10) |
| $\rho DF$ | 15.73 | (±17.97) | 3.48 | (±9.13) | 0.85 | (±1.21) | 7.25 | (±27.09) | 4.91 | (±18.90) | 4.05 | (±19.79) |
| combined | 21.66 | (±40.18) | 33.22 | (±220.89) | 1.14 | (±1.81) | 16.75 | (±67.43) | 24.13 | (±87.03) | 33.70 | (±220.28) |
| **Query class entity-o with 98 queries** | | | | | | | | | | | | |
| base | 2.81 | (±3.35) | 4.08 | (±4.89) | 1.3 | (±1.25) | 6.24 | (±5.52) | 1.39 | (±2.39) | 0 | (±0) |
| select | 2.8 | (±3.35) | 2.1 | (±2.83) | 1.24 | (±1.42) | 2.29 | (±3.75) | 0.31 | (±1.24) | 0 | (±0) |
| seeAlso | 2.8 | (±3.35) | 2.11 | (±2.71) | 1.2 | (±1.45) | 2.46 | (±3.91) | 0.36 | (±1.28) | 0 | (±0) |
| sameAs | 2.85 | (±3.42) | 3.38 | (±12.3) | 1.3 | (±1.44) | 2.4 | (±3.78) | 0.94 | (±3.73) | 3.13 | (±0.02) |
| $\rho DF$ | 2.86 | (±3.35) | 2.49 | (±4.39) | 1.56 | (±3.51) | 2.29 | (±3.75) | 0.65 | (±2.28) | 0.42 | (±1.42) |
| combined | 3.02 | (±3.54) | 2.71 | (±4.88) | 1.74 | (±4.15) | 2.57 | (±3.93) | 2.39 | (±7.18) | 4.61 | (±1.25) |
| **Query class entity-so with 70 queries** | | | | | | | | | | | | |
| base | 20.19 | (±34.99) | 7.83 | (±11.64) | 1.52 | (±4.4) | 16.9 | (±8.86) | 1.60 | (±2.62) | 0 | (±0) |
| select | 20.07 | (±35) | 2.9 | (±3.94) | 1.15 | (±1.71) | 5.09 | (±5.41) | 0.43 | (±0.53) | 0 | (±0) |
| seeAlso | 20.07 | (±35) | 2.56 | (±2.61) | 1.02 | (±1.21) | 5.16 | (±5.57) | 0.47 | (±0.60) | 0 | (±0) |
| sameAs | 24.53 | (±48.76) | 5.25 | (±17.71) | 1.04 | (±1.16) | 9.14 | (±24.22) | 3.57 | (±11.34) | 3.84 | (±21.65) |
| $\rho DF$ | 33.99 | (±56.32) | 3.44 | (±4.58) | 1.31 | (±2.96) | 7.3 | (±7.14) | 1.64 | (±2.03) | 1.18 | (±1.62) |
| combined | 45.66 | (±86.07) | 36.04 | (±271.09) | 1.34 | (±2.45) | 11.37 | (±25.25) | 11.61 | (±44.43) | 41.74 | (±32.78) |
| **Query class s-path-2 with 66 queries** | | | | | | | | | | | | |
| base | 1.76 | (±1.83) | 2.49 | (±2) | 1.15 | (±1.71) | 5.27 | (±2.33) | 0.90 | (±0.66) | 0 | (±0) |
| select | 1.76 | (±1.83) | 1.84 | (±1.74) | 0.96 | (±1.1) | 2.32 | (±1.95) | 0.26 | (±0.54) | 0 | (±0) |
| seeAlso | 1.76 | (±1.83) | 2.09 | (±2.6) | 1.05 | (±1.39) | 2.71 | (±3.06) | 0.30 | (±0.59) | 0 | (±0) |
| sameAs | 3.48 | (±8.26) | 17.41 | (±79.4) | 1.3 | (±1.59) | 6.62 | (±13.87) | 10.90 | (±46.04) | 27.41 | (±146.68) |
| $\rho DF$ | 2.59 | (±2.64) | 10.59 | (±22.48) | 5.44 | (±15.53) | 2.35 | (±1.94) | 0.61 | (±1.15) | 0.38 | (±0.68) |
| combined | 15.44 | (±95.93) | 17.89 | (±72.65) | 3.05 | (±11.43) | 6.95 | (±13.54) | 13.59 | (±51.97) | 34.84 | (±17.23) |
| **Query class o-path-2 with 56 queries** | | | | | | | | | | | | |
| base | 2.54 | (±9.58) | 6.21 | (±9.51) | 1.94 | (±2.02) | 7.02 | (±8.17) | 1.99 | (±3.60) | 0 | (±0) |
| select | 2.52 | (±3.99) | 2.55 | (±2.06) | 1.72 | (±1.74) | 2.55 | (±1.55) | 0.16 | (±0.25) | 0 | (±0) |
| seeAlso | 2.52 | (±3.99) | 2.87 | (±4.86) | 1.4 | (±1.94) | 3.16 | (±3.11) | 0.20 | (±0.41) | 0 | (±0) |
| sameAs | 2.77 | (±4.05) | 2.68 | (±2.34) | 1.74 | (±2.02) | 2.75 | (±1.76) | 0.52 | (±0.77) | 0.01 | (±0.06) |
| $\rho DF$ | 2.52 | (±3.99) | 2.55 | (±2) | 1.74 | (±1.53) | 2.55 | (±1.55) | 0.40 | (±0.53) | 0.24 | (±0.31) |
| combined | 2.77 | (±4.05) | 3.09 | (±4.92) | 1.64 | (±1.5) | 3.45 | (±4.43) | 1.98 | (±5.93) | 0.46 | (±1.59) |
| **Query class s-path-3 with 53 queries** | | | | | | | | | | | | |
| base | 1.87 | (±1.63) | 9.77 | (±18.93) | 2.14 | (±3.33) | 7.98 | (±3.96) | 3.04 | (±6.67) | 0 | (±0) |
| select | 1.87 | (±1.63) | 5.65 | (±6.83) | 2.13 | (±3.13) | 4.11 | (±3.29) | 2.26 | (±6.52) | 0 | (±0) |
| seeAlso | 1.87 | (±1.63) | 6.77 | (±12.84) | 1.93 | (±3.14) | 4.3 | (±3.24) | 2.29 | (±6.51) | 0 | (±0) |
| sameAs | 1.91 | (±1.63) | 22.79 | (±30.3) | 9.13 | (±12.48) | 4.55 | (±3.64) | 7.05 | (±9.19) | 0.06 | (±0.25) |
| $\rho DF$ | 2.6 | (±2.28) | 5.33 | (±7.15) | 1.93 | (±2.88) | 4.19 | (±3.28) | 4.19 | (±11.59) | 2.02 | (±5.28) |
| combined | 2.66 | (±2.28) | 7.03 | (±13.98) | 1.77 | (±2.96) | 4.81 | (±3.68) | 13.21 | (±34.69) | 2.13 | (±5.17) |
| **Query class o-path-3 with 45 queries** | | | | | | | | | | | | |
| base | 4.64 | (±11.94) | 10 | (±17.18) | 1.84 | (±3.12) | 11.53 | (±16.05) | 3.36 | (±4.77) | 0 | (±0) |
| select | 4.64 | (±11.94) | 6.51 | (±11.07) | 1.62 | (±1.18) | 5.42 | (±5.17) | 1.52 | (±3.39) | 0 | (±0) |
| seeAlso | 4.71 | (±11.93) | 6.76 | (±10.82) | 1.6 | (±0.99) | 6.44 | (±5.82) | 1.63 | (±3.44) | 0 | (±0) |
| sameAs | 6.13 | (±15.41) | 7.25 | (±11.64) | 1.7 | (±0.96) | 5.89 | (±5.28) | 5.13 | (±10.26) | 0.15 | (±0.49) |
| $\rho DF$ | 4.64 | (±11.94) | 6.73 | (±11.18) | 1.49 | (±0.84) | 5.42 | (±5.17) | 3.14 | (±7.10) | 2.85 | (±7.20) |
| combined | 6.18 | (±15.4) | 7.26 | (±10.29) | 1.74 | (±2.5) | 6.78 | (±6.5) | 11.11 | (±23.15) | 3.99 | (±5.20) |
| **Query class star-s3 with 91 queries** | | | | | | | | | | | | |
| base | 3.45 | (±13.13) | 4.05 | (±5.99) | 1.45 | (±3.91) | 7.66 | (±6.75) | 1.96 | (±7.58) | 0 | (±0) |
| select | 3.45 | (±13.13) | 2.68 | (±6.13) | 2.36 | (±6.08) | 1.35 | (±0.5) | 0.35 | (±1.52) | 0 | (±0) |
| seeAlso | 3.45 | (±13.13) | 2 | (±3.54) | 1.63 | (±3.51) | 1.43 | (±0.6) | 0.37 | (±1.52) | 0 | (±0) |
| sameAs | 3.57 | (±13.15) | 3.92 | (±16.77) | 1.79 | (±4.95) | 2.05 | (±5.76) | 2.55 | (±14.01) | 3.70 | (±34.99) |
| $\rho DF$ | 11.78 | (±75.27) | 2.24 | (±4.76) | 1.84 | (±4.61) | 1.35 | (±0.5) | 0.78 | (±3.03) | 0.43 | (±1.51) |
| combined | 11.97 | (±75.27) | 3.63 | (±14.92) | 1.69 | (±4.26) | 2.11 | (±5.76) | 3.84 | (±15.33) | 4.93 | (±42.49) |
| **Query class star-s2-o1 with 80 queries** | | | | | | | | | | | | |
| base | 1.71 | (±3.16) | 4.25 | (±5.85) | 2 | (±3.01) | 6.98 | (±5.77) | 1.41 | (±3.10) | 0 | (±0) |
| select | 1.71 | (±3.16) | 2.52 | (±6.53) | 2.21 | (±6.47) | 1.58 | (±0.59) | 0.14 | (±0.31) | 0 | (±0) |
| seeAlso | 1.71 | (±3.16) | 1.99 | (±2.54) | 1.62 | (±2.53) | 1.71 | (±0.77) | 0.14 | (±0.31) | 0.01 | (±0.01) |
| sameAs | 1.96 | (±3.24) | 2.98 | (±5.64) | 2.14 | (±4.98) | 2.49 | (±4.52) | 1.67 | (±7.47) | 1.27 | (±7.84) |
| $\rho DF$ | 2.66 | (±3.69) | 2.3 | (±3.81) | 1.95 | (±3.77) | 1.58 | (±0.59) | 0.35 | (±0.60) | 0.22 | (±0.32) |
| combined | 3.21 | (±4.44) | 2.48 | (±3.24) | 1.68 | (±2.19) | 2.61 | (±4.63) | 2.59 | (±9.11) | 1.74 | (±9.22) |
| **Query class star-s1-o1 with 69 queries** | | | | | | | | | | | | |
| base | 1.55 | (±4.21) | 2.85 | (±2.75) | 1.18 | (±1.4) | 5.67 | (±4.83) | 1.21 | (±2.61) | 0 | (±0) |
| select | 1.55 | (±4.21) | 1.43 | (±1.67) | 1.08 | (±1.39) | 1.39 | (±0.49) | 0.45 | (±2.58) | 0 | (±0) |
| seeAlso | 1.55 | (±4.21) | 1.84 | (±2.86) | 1.4 | (±2.75) | 1.52 | (±0.63) | 0.46 | (±2.58) | 0 | (±0) |
| sameAs | 1.55 | (±4.21) | 2.05 | (±3.93) | 1.65 | (±3.8) | 1.51 | (±1.04) | 1.35 | (±7.75) | 0.01 | (±0.02) |
| $\rho DF$ | 1.72 | (±4.31) | 2.83 | (±7.46) | 2.28 | (±7.3) | 1.39 | (±0.49) | 1.13 | (±4.89) | 0.71 | (±2.64) |
| combined | 1.72 | (±4.31) | 2.5 | (±5.63) | 2.01 | (±5.45) | 1.64 | (±1.1) | 3.49 | (±14.70) | 0.72 | (±2.60) |
| **Query class star-s1-o2 with 70 queries** | | | | | | | | | | | | |
| base | 1.77 | (±5.27) | 3.91 | (±5.76) | 1.75 | (±2.37) | 6.69 | (±1.72) | 1.40 | (±1.68) | 0 | (±0) |
| select | 1.77 | (±5.27) | 2.35 | (±3.99) | 2.03 | (±3.97) | 1.67 | (±0.5) | 0.1 | (±0.11) | 0 | (±0) |
| seeAlso | 1.77 | (±5.27) | 1.7 | (±1.46) | 1.37 | (±1.45) | 1.8 | (±0.65) | 0.1 | (±0.11) | 0 | (±0) |
| sameAs | 2.4 | (±7.13) | 2.58 | (±6.21) | 1.93 | (±5.6) | 2.14 | (±3.66) | 1.15 | (±7.12) | 0.87 | (±7.21) |
| $\rho DF$ | 1.86 | (±5.3) | 1.85 | (±1.91) | 1.52 | (±1.83) | 1.67 | (±0.5) | 0.24 | (±0.75) | 0.17 | (±0.13) |
| combined | 2.6 | (±7.24) | 1.99 | (±2.77) | 1.34 | (±1.12) | 2.26 | (±3.78) | 1.83 | (±8.74) | 1.14 | (±8.12) |
| **Query class star-o3 with 71 queries** | | | | | | | | | | | | |
| base | 2.56 | (±4.54) | 4.89 | (±5.3) | 2.25 | (±1.93) | 8.48 | (±7.97) | 2.12 | (±3.47) | 0 | (±0) |
| select | 2.52 | (±4.54) | 2.49 | (±2.04) | 2.2 | (±1.99) | 1.89 | (±0.32) | 0.16 | (±0.23) | 0 | (±0) |
| seeAlso | 2.52 | (±4.54) | 2.79 | (±2.82) | 2.36 | (±2.84) | 2.17 | (±0.76) | 0.16 | (±0.23) | 0 | (±0) |
| sameAs | 3.56 | (±9.4) | 2.66 | (±3.28) | 1.86 | (±1.73) | 2.51 | (±3.6) | 1.30 | (±7.14) | 0.88 | (±7.16) |
| $\rho DF$ | 2.52 | (±4.54) | 3.11 | (±4.25) | 2.77 | (±4.15) | 1.89 | (±0.32) | 0.29 | (±0.55) | 0.19 | (±0.38) |
| combined | 3.56 | (±9.4) | 3.24 | (±4.67) | 2.47 | (±4.05) | 2.73 | (±3.86) | 2.11 | (±8.93) | 1.20 | (±8.07) |

Table 3: Benchmark results with mean and standard deviation.